\documentclass[useAMS,usenatbib]{mn2e}
\usepackage{amsmath}
\usepackage{amssymb}
\usepackage{graphicx}
\usepackage{epsfig}
\usepackage{url}

\title[Skies And Universes]{Skies And Universes: Accessing  cosmological simulations and theoretical predictions.} 
\author[Klypin et al.]{A. Klypin$^1$, F. Prada$^2$,  and
J. Comparat$^3$\\
 \vspace{-0.2cm}\\
$^{1}$ Astronomy Department, New Mexico State University, Las Cruces, NM, USA\\
$^2$ Instituto de Astrof\'{\i}sica de Andaluc\'{\i}a (CSIC), Glorieta de 
     la Astronom\'{\i}a, E-18080 Granada, Spain \\
$^3$ Max-Planck-Institut f\"ur extraterrestrische Physik (MPE), Giessenbachstrasse 1, D-85748 Garching bei M\"unchen, Germany
\\
}
\begin{document}
\maketitle

\begin{abstract} Numerical simulations play a key important role in modern
  cosmology. Examples are plenty including the cosmic web -
  large scale structure of the distribution of galaxies in space -
  which was first observed in $N$-body simulations and later discovered in
  observations. The cuspy dark matter
  halo profiles, the overabundance of satellites, the Too-Big-Too-Fail
  problem are other examples of theoretical predictions that have a
  dramatic impact on recent developments in cosmology and galaxy formation. Large
  observational surveys such as e.g. SDSS, Euclid, and LSST are intimately
  connected with extensive cosmological simulations that provide
  statistical errors and tests for systematics.  Accurate predictions
  for baryonic acoustic oscillations and redshift space distortions
  from high-resolution and large-volume cosmological simulations are
  required for interpretation of these large-scale galaxy/qso surveys. However,
  most of the results from extensive computer simulations, that would
  be greatly beneficial if publicly available, are still in hands of
  few research groups. Even when the simulation data is available, sharing vast
  amounts of data can be overwhelming. We argue that there is an
  effective and simple path to expand the data access and
  dissemination of numerous results from different cosmological
  models. Here we demonstrate that public access can be effectively
  provided with relatively modest resources. Among different results,
  we release for the astronomical community terabytes of raw data of the
  popular Bolshoi and MultiDark simulations. We also provide numerous
  results that are focused on mimicking observational data and galaxy surveys
  for major projects. {\it Skies and Universes} is a community effort: data are
  produced and shared by many research groups.  We offer to other
  cosmologists and astronomers to host their data products in the
  \url{skiesanduniverses.org} space.
\end{abstract}
\begin{keywords}methods - cosmology  - survey 
\end{keywords}
\maketitle

\section{Introduction}
\label{sec:introduction}

Modern cosmology is a diverse science that to a large degree has changed
our views on the Universe.  Numerical simulations always played and
still play a special role in cosmology. The fact that the large-scale
distribution of mass is a complex cosmic web -- a system of filaments
connecting large clusters of galaxies -- was first revealed by
numerical simulations \citep{Doroshkevich1980,Klypin1983} and later
was discovered in observations \citep{deLapparent}. $N$-body
simulations of the evolution of primordial fluctuations made a number
of startling predictions including cuspy dark matter profiles
\citep{Flores1994,Moore1994,NFW1997} in collapsed halos and diverging numbers
of small dark matter satellites orbiting inside galaxies like our
Milky Way \citep{Klypin1999,Moore1999}.

For most of the time the results of large cosmological simulations
were available only to a small number of research groups that invested
substantial resources in code development and had access to
supercomputers. There were no motivations or even means to make the
results available for everybody. Dynamics of the field with the fast
progress in methods and computer hardware was also a factor impeding
dissemination of the computational data: results were becoming
outdated way too fast. However, the situation is changing because of a number of reasons. One
is the quality of simulations. At some moment the simulations became
so good that they stay very useful for a very long time. Examples
include the Millennium \citep{Millennium}, Bolshoi\citep{Bolshoi} and MultiDark DR1 \citep{Prada12} simulations that are
used for a cutting-edge research even after many years since they were
made. In turn, the long life of simulations relieves the pressure from
research groups that made the simulations to keep them for
themselves. Another reason is the recent trend of shifting the
emphasis from making simulations (and reaping benefits of first
analysis) to more diverse analysis of the existing simulations. Yet
another reason is pure technological: the cost of hard drives and
speed of internet connections. The cost of disk storage has decreased
to a level when even a relatively modest investment into hardware
allows small research groups to store vast amounts of data. Fast
internet connection potentially allows others to access the data.  To
summarise, we are in a stage when vast amounts of data can be made
available to the whole astronomical community.

The field of dissemination of results of cosmological simulations was
pioneered by two websites/projects:
Millennium\footnote{http://gavo.mpa-garching.mpg.de/Millennium/},
\citep{Lemson2006}) and MultiDark\footnote{multidark.org, now
  incorporated into \url{cosmosim.org}}, \citep{Riebe}). They made
publicly available some data, but the access was quite limited by the
scope of data (only halo catalogs and semi-analytical models are
avaiable) and by the access method. Both websites allow only SQL
queries and do not allow downloading of the data files.  This type of
data access has its positive and negative sides. On the positive side,
when accessing the data the user does not need to know how the data is
stored, and does not need to use any programming tools. This makes the
access simple. Yet, one cannot apply an elaborate analysis tools to
the whole computational box. It is very difficult to do it using SQL
queries.  Maintenance and adding new data to the database is labor
intensive process. As the result, this direction for dissemination is
still limited to few websites and they represent an important
limitation in the era of substantial demands for simulation data.

Allowing public access through direct downloading files is another
path, which seems to be promising, but had limited success so far.
The website of {\it DarkSkies}
simulations\footnote{http://darksky.slac.stanford.edu} released halo
catalogs and particles for $z=0$ snapshots. Halo catalogs and halo
merging trees for Bolshoi and some MultiDark simulations can also be
downloaded\footnote{http://hipacc.ucsc.edu/Bolshoi}. These are very
useful resources. The only problem is that there are few of them and
they are narrowly focused on some specific simulations and data
products.

The need for public access to large simulations is also driven by
existing and upcoming large observational projects such e.g. SDSS, DESI,
Euclid, LSST. These projects rely on theory when it comes to estimates
of error bars and systematics. In order to estimate the errors one
needs to have realistic ways of producing large mock galaxy catalogs
that mimic observational setup (e.g., boundaries and masks, completeness) and
systematic effects (e.g., selection functions and fiber collision).

In order to produce results that are close to what is observed,
simulations must undergo an extensive analysis, which at the end
produces what one may call ``products''. Mock galaxy catalogs is an
example of a popular product. Because the process of producing products can be
very complex, it is useful to provide public access to these products in an efficient platform.

In response to these demands -- both from the theory and observations
-- we developed tools and assembled large data sets to facilitate
public access to different types of products. These data are available on \url{skiesanduniverses.org}.

In Section \ref{sec:data} we list different available types of
data. In Section \ref{sec:analysis} we discuss tools to access and
analyse the data. Section \ref{sec:expand} presents ways of expanding the data sets.

\section{DATA}
\label{sec:data}
\subsection*{Raw data from $N$-body simulations}
There are different types of ``raw'' data available on
\url{skiesanduniverses.org}.  These include positions and velocities
of particles for many snapshots of Bolshoi, MultiDark, and GLAM
simulations:

\begin{itemize}
\item GLAM simulations \citep{GLAM} are done with a new Parallel PM code that uses a large
  homogenous mesh. As any PM simulation, GLAM simulations are mostly
  limited by available memory. GLAM simulations have force resolution
  ranging from 200 kpc to 1 Mpc. The number of particles is typically
  1-4 billion. We make available 10 realisations of these simulations
  with some simulations being available with many snapshots. In the near future we plan to release
  many more realisations.

\item Bolshoi \citep{Bolshoi,Klypin2016} simulations were done with the Adaptive Mesh Refinement
  (AMR) type code ART \citep{ART,ART2008}. Bolshoi simulations have 8 billion
  particles and reach the force resolution of 1~kpc. A large number of
  snapshots are released for the public. 

\item MultiDark simulations \citep{Klypin2016} were performed with a TREE code GADGET \citep{GADGET}. They
  have 56 billion particles and resolution 1-10 kpc. Because of the
  size of these simulations, only 1-4 snapshots are available for each
  of the MultiDark simulations. More will be made available in the near future.

\end{itemize}

A large number of halo catalogs for Bolshoi and MultiDark simulations
are available. 
The halos were identified with
spherical-overdensity halo finders Bound Density Maximum (BDM, \citet{Riebe}) and/or
Robust Overdensity Calculation using K-Space Topologically Adaptive
Refinement (RockStar, \citet{RockStar}).  BDM is described in Appendix A of Riebe et
al. (2011).  RockStar and BDM halo finders were extensively studied
and compared \citep{Knebe2011,Behroozi2013,RockStar}. Both BDM and RockStar identify
distinct halo and subhalos, but they use different algorithms to do
it. The main difference between BDM in RockStar is in the masses of subhalos with RockStar giving bigger masses. 
Circular velocities show much smaller differences.

Halo identification for GLAM simulations is done with a strip-down version of BDM halo finder. In this case only distinct halos are identified and listed in catalogs.

There are two types of BDM catalogs that differ by the definition of
the viral radius. Catalogs with filenames having capital letter V
(e.g., CatshortV.0416.DAT) are for the overdensity $360\rho_{back}$
(background density). Catalogs with capital letter W are for $200\rho_{crit}$
(critical density) for defining the halo boundary.

We also made available simulation data generated by other groups. At present the user can find halo catalogs for the Lomonosov suite of cosmological simulations \citet{Sergey}. Our site is offered to other groups who have the interest to make publicly available their data and products.

\subsection*{Processed data of $N$-body simulations}
There are a variety of products related with large cosmological
simulations available on {\it Skies and Universes}.  Halo catalogs are an
example of these products.  Other products include halo mass functions
and velocity functions measured by \citet{comparat_HMF_2017}.  We also
present nonlinear dark matter power spectra for $~15,000$ realisations
of GLAM simulations. Those can be used to study the power spectrum
covariance matrix -- the stepping stone for producing mock
catalogs. Dark matter density distribution functions for $~5000$
simulations with different resolution, filtering scales and box sizes
are given on {\it Skies and Universes}.

\subsection*{Mock galaxy catalogs}

Different types of mock galaxy/qso catalogs can be found on \url{skiesanduniverses.org}. We expect that more will be available in the near future for eBOSS, Euclid, DESI and eROSITA. Currently we provide: 
\begin{itemize}
\item Galaxy and QSO mock catalogs for the SDSS DR7
  \citep{Favole2017}, BOSS DR12 \citep{Sergio2016,Kitaura2016} and
  eBOSS  \citep{Favole2016,Sergio_QSOmock_2017}.
\item MultiDark-Galaxies Catalogs of galaxies made with the
  GALACTICUS, SAG and SAGE semi-analytical models of galaxy formation
  \citep{Knebe2017}
\item MultiDarkLens provides to the scientific community high quality
  weak lensing data using the MultiDark cosmological N-body
  simulations \citep{Carlo}
\item LoRCA \citep{Comparat2016} Galaxy mock catalogs for the Calar Alto Local Universe survey.

\end{itemize}

\subsection*{Observational products}

We also host galaxy spectra as well as catalogs of measured galaxy emission lines that constitute the basis to two studies of the emission line luminosity functions \citep[see][]{Johan_cosmos_2015,Johan_DEEP2VVDS_2016}. These are available with the pull down menu under Products, Observations. 

\section{Analysis}
\label{sec:analysis}

The raw particle data are written as unformatted fortran
files. Routines to read the data are provided and description of data
formats are given. These routines should be used as templates for
building analysis tools. The particle data sets are so large that one
should use parallel programming tools -- either MPI or OpenMP -- to
analyse the data. The data format used to store the data was developed
to facilitate the parallel processing. Data were split into a large
number of boxes covering the whole computational domain. Each box is
surrounded with $\sim 5$~Mpc-wide buffer allowing users to analyse
each box separately and then combine the results to produce final
result for the whole simulation. This splitting of the simulation was
also designed to make MPI parallelization easier.
Other data products are written as compressed ASCII tables.

Users should be aware about the size of available data
sets. Downloading of these large files may take days.

We do not provide support with data analysis. People,
who access the data in \url{skiesanduniverses.org} are supposed to
download the data they need and to build their own tools to analyse
the data. However, there are some codes that we plan to make available
to facilitate reading e.g. the MultiDark-Galaxy data and also perform part of the analysis.

\section{Expanding the net}
\label{sec:expand}

The main goal of the {\it Skies and Universes} space is to provide access
for the astronomical community with results from cosmological
simulations. These results are useful on their own to understand
processes related with the non-linear evolution of cosmic
structures. Some other products are specifically focused on large
survey observational projects. To facilitate the interaction between
the theory and observations we also provide links to recent
observational projects.

This data repository is not designed for an easy and simple access to
a small portion of large data sets. There are other sites, that
provide this type of access. Data on our site are for analysis that
requires extensive computer resources. Users are expected to carefully
choose the data they need and download them once.

While the amount and diversity of data on \url{skiesanduniverses.org}
are substantial, it is important to emphasise that this is just a
proof of concept to demonstrate that it is possible to provide access to the general
astronomical community to a vast amount of data having only very limited
computer resources. This is the reason why we stay away from creating
traditional databases, which are used in cosmosim.org and millennium
database. 

It is relatively straightforward to expand the data available through
\url{skiesanduniverses.org}. We can host significant data sets. Larger
datasets can be made available by cross-linking data visible on the
internet. 

We encourage other groups to contact us if they are interested to 
host their simulation products in \url{skiesanduniverses.org}.

\section*{Acknowledgements}
\vspace{0.2cm}

We thank New Mexico State University (USA) and Instituto de
Astrof\'isica de Andaluc\'ia CSIC (Spain) for hosting the Skies and
Universes site for cosmological simulation products. 
We also thank the Instituto de Fisica Teorica (IFT/CSIC-UAM) for 
hosting the first version of our site. F.P. acknowledges support from the 
Spanish MINECO grant AYA2014-60641-C2-1-P.

Supercomputing resources are vital for producing the data posted
here. We acknowledge support from NASA Ames Research Center (USA),
Leibniz-Rechenzentrum (LRZ, Germany), Barcelona Supercomputer Center
(Spain) and CESGA (Galicia, Spain). 

We want thank the main contributors that deliver the data hosted in the  \url{skiesanduniverses.org} website:
\begin{itemize}
\item Bolshoi simulations: Anatoly Klypin (NMSU), Joel Primack (UCSC)
\item GLAM simulations: Anatoly Klypin (NMSU), Francisco Prada (IAA-CSIC)
\item MultiDark simulations: Gustavo Yepes (UAM), Stefan Gottloeber (AIP), Anatoly Klypin (NMSU), Francisco Prada (IAA-CSIC)
\item Lomonosov simulations: Sergey Pilipenko (Lebedev)
\item BDM halos: Anatoly Klypin (NMSU)
\item RockStar halos: Peter Behroozi (Berkeley)
\item MultiDark-Galaxies: Alexander Knebe (UAM), Doris Stoppacher (IFT-UAM/CSIC), Francisco Prada (IAA-CSIC), Andrew Benson (Carnegie), Sofia Cora (La Plata), Darren Croton (Swinburne)
\item MultiDarkLens: Eric Jullo (LAM), Ben Metcalf (Bologna), Carlo Giocoli (Bologna), Sylvain de la Torre (LAM)
\item SDSS galaxy and QSO mock catalogs: Sergio Rodriguez-Torres (UAM), Johan Comparat (MPE), Ginevra Favole (ESAC), Francisco Prada (IAA-CSIC), F. Kitraura (AIP).
\item LoRCA, observational products, analytical approximations: Johan Comparat (MPE)
\end{itemize}

\bibliographystyle{mn2e}
\bibliography{skiesanduniverses}

\end{document}